\documentstyle[aps,12pt,epsfig]{revtex}
\begin{document}
\draft
\def\seteps#1#2#3#4{\vskip#3\relax\noindent\hskip#1\relax \special{eps:#4 x=#2, y=#3}}
\def\centereps#1#2#3{\vskip#2\relax\centerline{\hbox to#1{\special   {eps:#3 x=#1, y=#2}\hfil}}}

\def\NI{\noindent} \def\lss{L. S. Schulman}
\def\ssc{\scriptscriptstyle } 
\def\Tr{\mathop{\rm Tr}\nolimits}
\def\note#1{\def\dash{\hbox{\rm---}}{\bf~[[}~{\tt #1}~{\bf]]~}}
\def\doublelinebreak{\medskip \hrule \vskip 2pt \hrule \medskip}
\def\continue{\bigskip\bigskip\note{continue$\dots$} \doublelinebreak\bigskip}

\title{Resolution of causal paradoxes arising from opposing thermodynamic arrows of time} 
\author{L. S. Schulman}
\address{Physics Department, Clarkson University, Potsdam, New York
13699-5820}
\date{\today}
\maketitle
\begin{abstract}
It was recently shown \cite{opposite} that systems with opposite thermodynamic arrows of time could have moderate mutual interaction with neither destroying the order of the other. Such interaction includes signaling. Signals, however, may give rise to causal paradoxes, suggesting that ``moderate" interaction may be impossible. Using the two-time boundary condition framework, we resolve the paradoxes. In one example, at a macroscopic level, we establish the existence of solutions to the appropriate boundary value problem. This result is extended to a class of microscopic problems. We also produce an example in which microscopic data are given and there is no solution. This is a different kind of resolution: there is no paradox because the events do not happen. Finally we discuss the differences between these cases.
\end{abstract}



\section{Introduction} \label{intro}

The use of macroscopic {\it initial\/} conditions is equivalent to the imposition of a particular thermodynamic arrow of time \cite{timebook}. A non-circular study \cite{correlating} of the origins of an arrow, or of the simultaneous coexistence of two opposing arrows \cite{opposite}, therefore requires a more general formulation, and we have used the following scheme: give data at the two endpoints of the time interval within which the arrow is to be studied. For such two-time boundary conditions one gives less information at each end than would be appropriate for, say, initial conditions (this avoids overdetermination).

Using this framework, we have argued \cite{timebook,correlating} for Gold's thesis \cite{gold} relating the thermodynamic and cosmological arrows of time. More recently we showed that systems with opposite running arrows could coexist without either destroying the order of the other \cite{opposite}. Using an extension of the Wheeler-Feynman absorber theory \cite{absorberA,absorberB}, it was also demonstrated that electromagnetic signals can pass between the systems. The feasibility of other kinds of signaling was shown in \cite{oppvail}.

Within the opposing-arrow framework causal paradoxes occur, and in \cite{opposite} no resolution was offered. This is a crucial issue, since those paradoxes have been used \cite{penrose} to argue against Gold's thesis. Furthermore, if the passage of signals (weak in energy, say, but potentially strong in their effect on the systems) were forbidden under the weak coupling assumption, then the observability of opposite-arrow regions would be called into question.

The key idea for the resolution of the paradoxes is the precise posing of the problem, and we begin by presenting a typical paradox.

\section{The carpet paradox}\label{carpet}

In its narrative form \cite{opposite} the causal paradox is phrased in terms of system A (Alice) and system B (Bob). They have opposite thermodynamic arrows. See Fig.~1. At her 8 a.m., Alice sees rain coming in Bob's open window, ruining his carpet. It is 5 p.m.\ his time. At 8:20 a.m.\ (her time), Alice sees the rain begin near Bob. At 9 a.m.\ she sends a message, ``It will rain. Close your window." Bob receives this at 4 p.m., his time (40 min.\ {\it ``before"} the rain starts), and shuts the window. What did Alice see?

For this paradox to represent a bona fide inconsistency {\it one needs a set of boundary conditions at the remote times, conditions that lead to the indicated events}.

Phrasing the problem as we have in the last sentence shows that in a sense there is already a trivial resolution. If one poses a boundary value problem without solution, then these events simply do not happen. Indeed in a later section we will give an example of a two-time boundary value problem, suitable for the two-arrow framework, without solution. 

What is interesting is that if one assumes {\it continuity} in nature and compactness of some of the available state spaces, then boundary conditions that lead to apparent inconsistency in the A-B-carpet story, do in fact have a solution.

\subsection{Boundary value formulation}

We rephrase that story as a two-time boundary value problem. There are two two-state systems. The first is Alice. In one state, call it ``0," she does not signal; in the other state (``1") she does. Her initial condition (7:45 a.m., her clock) is $A(\hbox{7:45 a.m.})=0$: she will not signal unless she sees the need. The other two-state system is Bob's window. Define this to be ``0" if the window is closed, ``1" if open. Bob's ``initial" condition---3:45 p.m., his clock---is open, $B(\hbox{3:45 p.m.})=1$.

Let $\alpha \equiv A(\hbox{8:40 a.m.})$, Alice's intended message at 8:40 a.m. Take $\beta \equiv B(\hbox{4:20 p.m.})$, the state of Bob's window at 4:20 p.m. Consider the function $\beta=\psi(\alpha)$. If $\alpha=1$, Alice sends the message, Bob closes his window, and $\beta=0$. If $\alpha=0$, she does not send and (because of Bob's ``initial" condition) $\beta=1$. Thus $\psi$, as a function, reverses 0 and 1. Define a second function $\alpha=\phi(\beta)$ describing Alice's response to the state of Bob's window. If $\beta=1$, Alice will see rain entering Bob's window and her message will be to close the window ($\alpha=1$). Similarly if $\beta=0$ no message is necessary and $\alpha=0$. So $\phi$ is the identity.

The paradox arises by concatenating the functions, writing $\beta$ as a function of itself: $\beta=\psi(\alpha)=\psi(\phi(\beta))$. Using $\phi=$ identity, this implies $\beta=\psi(\beta)$. This is a contradiction, since $\psi$ reverses the two values in its domain and therefore has no fixed point.

\subsection{Resolution}

We resolve the paradox by appealing to continuity in nature and to the compactness of certain of the spaces of available states. Continuity should apply to finite systems with non-singular potentials. Similar appeals were made in \cite{absorberB} and \cite{tachyon}. The assignment of two-state variables to $A$ and $B$ is refined by giving them values in the interval $I\equiv[0,1]$, with the extreme values having their previous meanings. For example, $B$ no longer need represent a fully open or fully closed window. (We later discuss larger internal spaces.) Moreover, the maps should be continuous. $\phi$ as identity is replaced by a more realistic function $\phi:I\to I$. This continuous function, which by a mild abuse of notation we continue to call ``$\phi$," still takes the values 0 and 1 at the respective endpoints.  An example of an appropriate function with a steep threshold is 
$$
\phi(x)\equiv 
 \frac12 \left[ 1- \frac{\cot(\pi x)}{\varepsilon +|\cot(\pi x)|}\right] \,,
$$
and $\varepsilon\ll1$. Similarly, we can write $\psi=1-\tilde\phi$ for some interpolating function like $\phi$.

As before, concatenate the functions to get a function $\Psi: \{\beta\} \to \{\beta\}$, where now $\{\beta\}=I$. Again, $\Psi(0)=1$ and $\Psi(1)=0$. However, this $\Psi$ is continuous \cite{finepoint}. By general theorems it is known that $\Psi$ must have a fixed point within the interval. This fixed point yields a solution to the boundary value problem, hence a resolution of the paradox, as we momentarily explain.

For higher dimensional internal spaces a fixed point also exists, provided at least one of the spaces is compact (e.g., the states of the window) and the maps are continuous. This is because the concatenation becomes a mapping of this space into itself, and by standard topological theorems one is guaranteed the existence of a fixed point. See \cite{edwards,spanier}, and for a similar application, \cite{tachyontwo}.

What is the significance of the fixed point? It means that although we {\it tried\/} to set the boundary value problem as a contradiction, a solution exists, but it departs from a simple on/off description. Moreover, since all is determined by the boundary values, participants cannot ``change their minds." For the soggy carpet scenario, the solution {\it could\/} take the following form. Again, at 3:45 p.m.\ Bob's window is open and at 7:45 a.m.\ Alice does not plan to send a signal, but is disposed to help Bob if the need should arise. At 8 a.m.\ she sees his window slightly open and a bit of rain entering. She wonders if this will cause damage and vacillates. Ultimately she signals, but less stridently: perhaps, ``Some rain will enter; you might close your window." He reacts by balancing his desire for fresh air with her warning by closing the window partially. A self consistent solution has been found. Similar paradox resolutions have been used in science fiction concerned with time travel \cite{scifi}. Heinlein, in particular, presented many examples of self-consistent loops, generally of greater dramatic interest than that given above. Here is an observation of his on grammar, related to the need above to put quotation marks around certain words (e.g., ``before"):

\begin{quote}
Then I caught hold of myself and realized that, out of all the persons living in 1970, he was the one I had least need to worry about. Nothing could go wrong because nothing had $\dots$ I meant `nothing would.' No---Then I quit trying to phrase it, realizing that if time travel ever became widespread, English grammar was going to have to add a whole new set of tenses to describe reflexive situations $\dots$ (page 241, \cite{door}).
\end{quote}

\section{Generalized two-time mixed boundary value problems}\label{generalized}

In the previous section, we argued that the formulation of causal loop paradoxes, like the formulation of any physical problem in which a thermodynamic arrow of time is not presupposed, requires giving data at the ends of the time interval of interest. The peculiarity of the opposite arrow situation (in contrast to more general studies in which the entire history of the universe is contemplated) is that we focus only on a slice out of the universe's history. Moreover, it is assumed that the system under study breaks up naturally into two pieces. These have opposite thermodynamic arrows of time, imposed by the boundary conditions. The physical reason for the appropriateness of these boundary conditions could be differences in the individual histories of interaction or lack thereof with other parts of the cosmos.

In Sec.\ \ref{carpet} we used this formulation, rephrasing a paradox as a boundary value problem, with ``initial" conditions given for each subsystem---initial with respect to the clock of that subsystem. The natural next question is to go beyond the specific issue of paradoxes and ask whether this kind of mixed boundary value problem has or has not a solution in general.

\subsection{Oscillators: a microscopic example} \label{oscillators}

We begin with a simple example showing that the mixed boundary value problem need {\it not\/} have a solution (and will later explain how this differs from the carpet paradox).

Consider systems A and B, with $t=0$ conditions given for A and $t=T$ conditions for \hbox{B}. The example needs only a single one-dimensional particle in each system, particle \#1 in A, particle \#2 in B. Let the Hamiltonian be
$$ H=\frac12\left(p_1^2 + p_2^2 \right) + \frac12\rho\left(x_1-x_2\right)^2$$
with $\rho$ a real, positive parameter. The general solution is
$$x_{\left(2\atop1\right)}=X_0+V_0t\pm a\cos(\omega t-\phi)$$
with $\omega=\sqrt{2\rho}$ and $X_0$, $V_0$, $a$ and $\phi$ constants fixed by the boundary conditions. For simplicity we take $x_1(0)=0$ and $\dot x_1(0)=0$, which implies that $X_0=a\cos\phi$ and $V_0=a\omega\sin\phi$. The final boundary conditions take the form
\begin{eqnarray}
&x_2(T)=&
        (a\cos\phi)\left(1+\cos\omega T\right)
		 +(a\sin\phi)\left(\omega T +\sin\omega T\right)  \nonumber\\
&\dot x_2(T)=&(a\cos\phi)\left(-\omega\sin\omega T\right)
       +(a\sin\phi) \left(\omega + \omega\cos\omega T\right) \nonumber
\end{eqnarray}
It follows that (generically) there is no solution for the unknowns $a\cos\phi$ and $a\sin\phi$ if the determinant of the following matrix vanishes:
$$\pmatrix{1+\cos2\theta &2\theta +\sin2\theta\cr
           -\sin2\theta  & 1+\cos2\theta\cr}
$$
with $\theta=\omega T/2$. This happens whenever either $\cos\theta=0$ or $\theta\tan\theta=-1$. Therefore at those special times for which (at least) one of these equations is satisfied there exist final conditions on particle \#2 (system B) for which no solution to the mixed two-time boundary value problem exist.

One can extend this to 4 particles (A and B have 2 apiece). Let all have unit mass and let the potential be
$$ V= v(x_1,x_2)+ v(x_3,x_4) + \rho \left[v(x_1,x_3) + v(x_1,x_4) + v(x_2,x_3) +v(x_2,x_4)\right] $$
with $v(x,y)\equiv(x-y)^2/2$. The oscillation frequencies for this system are 0, $\omega\equiv\sqrt{2(1+\rho)}$ (twice), and $\omega'\equiv2\sqrt{\rho}$. Giving zero initial positions and velocities for the particles in A, the final boundary conditions in B are expressed linearly in terms of combinations of the normal mode parameters. Again there is a matrix whose determinant fixes the invertibility of this relation. That matrix turns out to be singular for exactly the conditions given for the two particle system, but with $\theta=\omega'T/2$.

\section{Which resolution---existence or non-existence?}

Despite our exhibiting two sorts of resolution---existence and non-existence---there is still a puzzle: for the carpet paradox we proved that there is a solution to any mixed boundary value problem of the sort considered; for the oscillator example we exhibited boundary values without solution. How can the ``carpet" proof be correct in the light of the oscillator example?

There are significant differences between these examples. The most obvious is the fact that in one case macroscopic reasoning is used; in the other detailed specification of the full microscopic boundary value problem is given. A second difference is that in the oscillator example the coordinate space is infinite, leading to yet another feature, unboundedness of the potential. The latter property can also arise in a finite coordinate system for an appropriate potential. Unbounded space or potential energy in turn can lead to non-compactness of the system phase space.

In the next subsection we show how changing some of these properties leads to existence proofs even with microscopic boundary conditions.

\subsection{Existence of a microscopic solution}

With assumptions that are reasonable for the carpet paradox, one can show existence, for sufficiently weak coupling, of a solution to a microscopic boundary value problem in which each subsystem, A and B, is given the correct amount of data to fully determine its motion in the absence of the other, but with the data for A given at $t=0$ and those for B at $t=T$. This will be called a mixed boundary value problem. The assumptions are these:

\smallskip

\NI 1.~ There is no exchange of particles. The subsystems retain their identities and the interaction is handled as a potential between particles.

\smallskip

\NI 2.~ The (classical) Hamiltonian can be written (in obvious notation) in the form $H= H_{\ssc A} + H_{\ssc B} + \lambda H_{\ssc AB}$, with $\lambda$ a real parameter.

\smallskip

\NI 3.~ For a fixed time interval ($[0,T]$) the final points (in phase space) are a continuous function of the initial points as well as of $\lambda$.

\smallskip

Note that it is not our objective to show that these assumptions must always be true, only to reconcile the ``carpet" resolution with the absence of solution in the oscillator example.

\def\gma{\gamma_{\ssc A}} \def\gmb{\gamma_{\ssc B}} 
\def\gmat#1{\gamma_{\ssc A}(#1)} \def\gmbt#1{\gamma_{\ssc B}(#1)} 
\def\gmatl#1#2{\gamma_{\ssc A}(#1;#2)} \def\gmbtl#1#2{\gamma_{\ssc B}(#1;#2)} 
\def\gmbzz{\bar\gmb^{{\ssc0}}(0)}
\def\nbhdinitial{{\cal N}_0}
\def\nbhdfinal{{\cal N}_F}

We introduce the following notation. Let $\gamma_{\ssc X}(t)$ be the phase space point for subsystem X (X is A or B) at time-$t$. The boundary conditions are $\gmat0=\bar\gma$ and $\gmbt T=\bar\gmb$. 
Our objective is to show that there exists $\gmbt0$ such that the (total system) phase space point $(\bar\gma,\gmbt0)$ evolves (for $t$ going from 0 to $T$) under the full time evolution (with non-zero $\lambda$) to the (total) system phase space point $(\gmat{T}, \bar\gmb)$ (for some $\gmat{T}$).

First consider the situation for $\lambda=0$. The individual phase space points, $\bar\gma$ and $\bar\gmb$ can be used to move forward or backward in A and B, separately.  We evolve the point $\bar\gmb$ back from time-$T$ to time-0, yielding a point that we designate $\gmbzz$. Let $\nbhdinitial$ be a neighborhood of $\gmbzz$. We proceed in several steps.

\smallskip

\NI 1.~ Evolve $\nbhdinitial$ forward in time, with some $\lambda\neq0$. For this evolution use $\bar\gma$ as the initial point for subsystem A. The time-$T$ image of $\nbhdinitial$ is an open set. Call it $\nbhdfinal(\lambda)$.

\smallskip

\NI 2.~ For any particular $\lambda$, the point $\bar\gmb$ may or may not be in $\nbhdfinal(\lambda)$. However, for $\lambda=0$, it is in the {\it interior\/} of $\nbhdfinal(0)$.

\smallskip

\NI 3.~ If, for a particular non-zero $\lambda$, $\bar\gmb \in \nbhdfinal(\lambda)$, then we are finished. For this means that there is a point in $\nbhdinitial$ that (together with $\bar\gma$) serves as an initial condition for reaching the desired final value of subsystem B.

\smallskip

\NI 4.~ Suppose then that for the $\lambda$ chosen, $\bar\gmb \notin \nbhdfinal(\lambda)$. Now decrease $\lambda$. Since $\bar\gmb$ is ultimately (i.e., for $\lambda=0$) in the interior of $\nbhdfinal$, it must enter and be in the interior for some finite $\lambda$, strictly greater than zero. For this value of $\lambda$ there is thus an initial condition in $\nbhdinitial$ for which subsystem B reaches the desired final state.

\subsection{Other differences between the examples}

Although there is no longer a conflict between the two cases studied, it is worth commenting on differences between them and on whether existence or non-existence of a solution should be expected in any particular situation. We list considerations that appear relevant.

\medskip

\NI 1.~ {\it Features of the microscopic dynamics}

\smallskip

\NI 1a.~
The infinite-range linear forces of the oscillator are an idealization. It was their harmonic forces that led to special times for which the boundary value problem could not be satisfied. As an instructive example, consider the usual, single system, two-time boundary value problem: $x(0)=0$, $x(2\pi)=a$. For a unit-mass particle in the potential $V(x)=\frac12x^2$ there is no solution for $a\neq0$. However, for small cubic or quartic contributions to $V$ there is always a solution for sufficiently small~$a$.

\smallskip

\NI 1b.~
Going beyond harmonic forces, a more interesting boundary value problem failure would occur if for sufficiently long times there were finite-measure regions of phase space for which unsolvable problems could be posed.

\smallskip

\NI 1c.~
The exhibited incompatible boundary conditions are not generic, in the sense that it is only for special times that the failure occurs.

\medskip

\NI 2.~ {\it Features of the macroscopic boundary value problem}

\smallskip

\NI 2a.~
Macroscopic data consist of coarse grains consistent with many microscopic states. Suppose that for realistic forces one had microscopic boundary conditions without solution on a strictly-positive-measure set of phase space (as contemplated in paragraph \#1b). If these regions were finely interspersed among coarse grains, with chaotic dynamics the macroscopic problem could still have a solution despite the existence of unsolvable microscopic specifications. For symmetric two-time boundary value problems this does happen (see \cite{timebook}, Sec.~5.1).

\smallskip

\NI 2b.~
The macroscopic arguments involve emergent concepts, features that can only be extracted from the microscopic state with difficulty. In particular, there is the notion of {\it signal}, used in an essential way in the carpet story.

\medskip

\NI 3.~ {\it Other considerations}

\smallskip

\NI 3a.~
Discussion until now has ignored the quantum nature of the world. However, in a microscopic version of the paradox one should use quantum mechanics. One way to do this was proposed by Gell-Mann and Hartle \cite{gellmann}. One calculates the probability of a sequence of events by sandwiching appropriate projection operators (for those events) between {\it two} density matrices, $\rho_i$ (initial) and $\rho_f$ (final). The normalization for this probability is $N\equiv 1/\Tr\rho_i\rho_f$. In their discussion, where the initial and final times are cosmologically separated, $N^{-1}$ will be {\it very} small. Nevertheless, it is not expected to be zero. The same phenomenon should provide solutions to two-time boundary conditions even where classically unavailable. 

It is easy to demonstrate this feature explicitly for our oscillator example. Two-time boundary conditions come naturally to quantum mechanics when the path integral formulation is used. For the oscillators, an analog of the classical problem is to give coherent state boundary conditions. Such states can be labeled by phase space points. Thus one sandwiches the propagator between initial and final $\gamma$s, in our earlier notation. All we must show then is that the sandwiched propagator is not zero at any time. For this purpose we can stay with the simpler center-of-mass coordinates, instead of going to $x_1$ and $x_2$ as done in Sec.~\ref{oscillators}. Now before integrating the propagator with the coherent state Gaussians it {\it does} vanish. Specifically, we recall the propagator for a mass-$m$, frequency $\omega$, oscillator with coordinate $\xi$ (the center-of-mass coordinate of our pair), to go from $\xi_1$ to $\xi_2$ in time $T$:
$$G(\xi_2,T;\xi_1)=\sqrt{\frac{m\omega}{2\pi i \sin\omega T}}
                   \exp\left\{\frac{im\omega}{2\sin\omega T}
						          \left[\left(\xi_1^2+\xi_2^2 \right)\cos\omega T
									 -2\xi_1\xi_2  \right] \right\}  \,.
$$
For $\omega T=n\pi$, with $n$ an integer, this is a delta-function; in particular it vanishes if given inappropriate endpoints. Now sandwich $G$ between coherent states, $|z_1\rangle$ and $|z_2\rangle$, where we use the usual notation. The resulting quantity, $\langle z_2|G(T)|z_1\rangle$ will equal $\langle z_2|z_1\rangle$ whenever $\omega T=2n\pi$ because $G$ is the identity operator. The quantity $\langle z_2|z_1\rangle$ never vanishes. (For odd-integer multiples of $\pi$ the matrix element will involve a reflected $z$, but again it never vanishes.) It follows that with the Gell-Mann-Hartle formulation of two-time boundary conditions, there will always be a state satisfying the quantum version of the oscillator boundary value problem given in Sec.~\ref{oscillators}.

\section{Conclusions}

We have found that purported causal paradoxes do not constitute a bar to logical consistency nor to the possible detection of opposite-arrow regions, should they exist. If one is confronted by a causal paradox within the context of a system possessing opposing arrows, the resolution of the paradox may allow a version of the scenario to be realized, or the resolution may be that there is no solution at all---the main point is that in all cases the paradox must be phrased as a mixed boundary value problem, and solutions sought, with no guarantee of finding them in all cases.

\acknowledgments
I am grateful to P. Facchi, E. Mihokova, D. Mozyrsky, C. M. Newman, S. Pascazio, A. Scardicchio and L. J. Schulman for useful discussions and comments. This work was supported in part by National Science Foundation grant PHY 9721459.


\def\sqmeasw{6 truein} \def\sqmeash{4 truein}
\def\precaption{.5 truein}

\begin{figure}[t]

\hbox{\hfill\epsfig{file=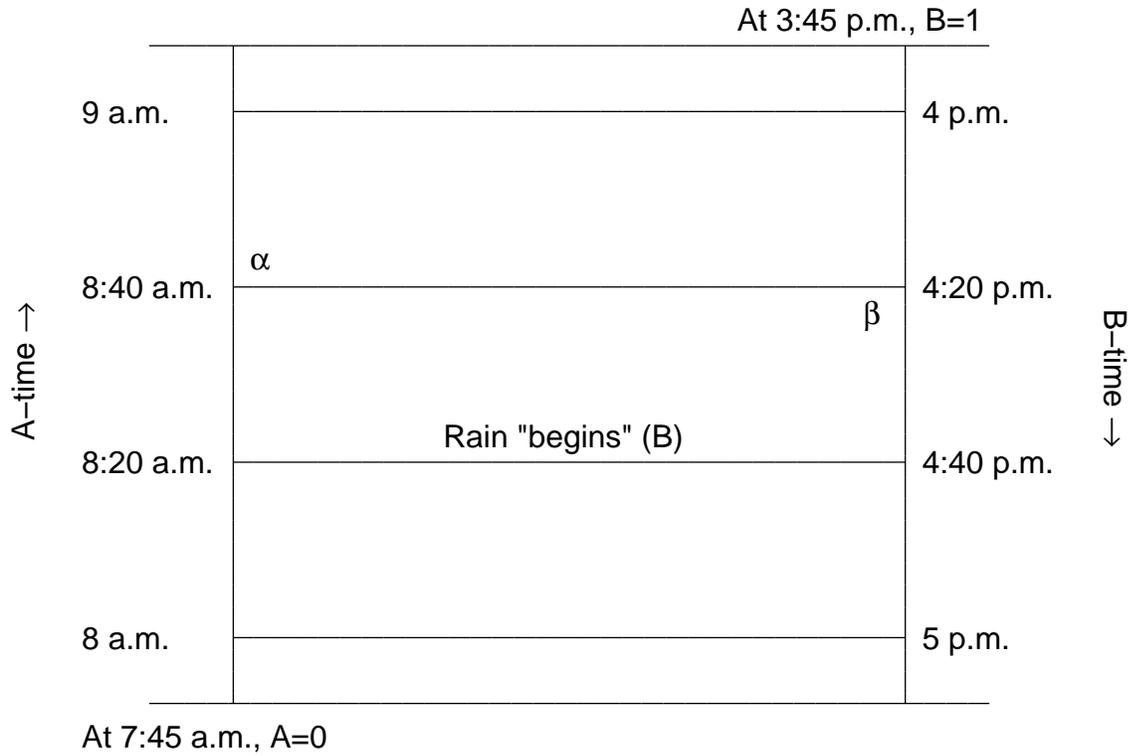,height=\sqmeash,width=\sqmeasw}}

\vskip \precaption
\caption{Times, variables and functions in the causal paradox. For the boundary value problem formulation of the paradox, each system is given initial conditions, where ``initial" is relative to that subsystem's clock. Thus at 7:45 a.m.\ her Alice does not ``yet" see the need to send a signal. At 3:45 p.m.\ his time, Bob's window is open.}
\label{fig1}
\end{figure}


\begin{references}
\bibitem{finepoint} Because the continuous extension, $\phi$, is no longer trivial, the concatenated function, $\Psi$, is not the same as $\psi$, the continuous extension of the discrete $\psi$.
\end{references}
\end{document}